# Development of a cavity ring-down spectroscopy sensor for boron nitride sputter erosion in Hall thrusters

L Tao[1], N Yamamoto[2], A D Gallimore[3] and A P Yalin[1]

1 Department of Mechanical Engineering, Colorado State University, Fort Collin, CO, 80523, USA

2 Department of Advanced Energy Engineering Science, Kyushu University, Kasuga, Fukuoka, 8168580, Japan

3 Department of Aerospace Engineering, University of Michigan, Ann Arbor, Michigan, 48109, USA

E-mail: azer.yalin@colostate.edu

**Abstract.** Sputter erosion of boron nitride (BN) is a critically important process in Hall thrusters from the point of view of both lifetime assessment and contamination effects. This contribution describes the development of a laser based sensor for *in situ* monitoring of sputtered BN from Hall thrusters. We present a continuous-wave cavity ring-down spectroscopy (cw-CRDS) system and its demonstrative measurement results from BN sputtering experiments.

## 1. Introduction

Hall thrusters are a type of electric propulsion (EP) engine used in space propulsion applications. They were developed in the 1960s to alleviate the thrust density limitations and large grid erosion rates of gridded ion thrusters. Hall thrusters offer an attractive combination of high thrust efficiency ( >50%), specific impulse of 1000–4000 s, and high thrust density allowing compact propulsion design [1-3]. These parameters make Hall thrusters suitable for space exploration missions, as well as satellite station keeping and orbit transfer. Since their first application in 1971 [4], over 200 Hall thrusters have been operated in space. Typical mission profiles require long operation lifetimes, for example a Hall thruster used for north-south station keeping (NSSK) of a commercial spacecraft will have to operate for over 5,000 hours over the course of its mission [5,6].

The primary mechanism limiting the lifetime of Hall thrusters is sputter erosion of the acceleration channel wall that protects the magnetic circuitry from the discharge plasma [2, 5-7]. Not only does sputter erosion end the life of the Hall thruster once it etches though the insulator wall material, but it also causes redeposition of sputtered wall material forming unwanted coatings on spacecraft surfaces such as solar arrays and sensor optics, thereby posing a serious risk to spacecraft operation. Boron nitride (BN) is the most widely used material for the acceleration channel wall in Hall thrusters. Currently, the primary way of verifying that a Hall thruster has sufficient life for its mission is to operate it beyond its expected total thrust duration in a vacuum chamber, a procedure referred to as ground-based life testing [5-7]. With proposed thruster missions now as long as 5-10+ years, ground-based life tests are becoming increasingly impractical and expensive. In addition to high cost and time intensiveness, one is limited to *post facto* analysis which does not readily allow accelerated measurements or study of varying operating conditions. Thus, it is very desirable to develop advanced diagnostics capable of *in situ* thruster measurements. The ideal diagnostic should have high sensitivity to measure low erosion rates, the possibility of integration to a thruster test facility, and fast time response to explore a range of operating conditions. The need for a sensitive non-intrusive measurement suggests the use of optical techniques. Optical emission spectroscopy (OES) [8,9], laser induced fluorescence (LIF) [10-12], and multi-photon ionization coupled to a time of flight mass spectrometer [13-15] have all been used for species-specific sputter measurements. The use of LIF has been particularly extensive, and has proven to be very effective for velocity measurement though challenging for quantitative number density. OES spectroscopy is attractive owing to its experimental simplicity but the analysis can be challenging since collisional-radiative modeling is required to extract overall species concentrations from the measured excited state number densities. OES has also been used for in situ thruster erosion measurements but again requiring rate assumptions or modeling [16-18].

This contribution focuses on developing a cavity ring-down spectroscopy (CRDS) sensor to study sputter erosion of boron nitride. CRDS is a highly sensitive laser-based absorption method allowing species-specific measurement of concentrations of trace species. The technique is used extensively for quantitative trace-species measurement in flames, plasmas, and the atmosphere [19,20]. Our previous research efforts have demonstrated its use for the study of sputtered particles in various configurations [21-28]. The possibility of a near real-time sputter sensor was shown for an industrial ion beam etch system [25]. A CRDS sensor detecting sputtered manganese for erosion studies of an anode layer type thruster has also been recently demonstrated [26]. Our past work has additionally shown detection of sputtered BN using pulsed and continuous-wave laser [27,28]. Here we describe the development of a CRDS sensor for BN erosion from Hall thrusters. The layout of the paper is as follows. Section 2 describes the diagnostic technique and optical detection scheme. Section 3 presents the experimental setup including the sputter facility and CRDS sensor. Demonstrative results and validation as well as sensitivity analysis are considered in Section 4, and conclusions are presented in Section 5.

**2. Cavity Ring-Down Spectroscopy**

*2.1. Technique Overview*

CRDS is a direct absorption spectroscopy measurement technique that provides extremely high sensitivity by employing enhanced optical-path length. The basic idea of CRDS is to measure the absorption spectrum of a sample (e.g. collection of sputtered particles) that is housed within a high-finesse optical cavity, typically formed from a pair of high-reflectivity mirrors. The laser beam is injected into the optical cavity where it is reflected back and forth many times, e.g. $\sim 10^4$ passes for a mirror reflectivity of $R \approx 0.9999$. The intensity of trapped light inside the cavity decays exponentially with time [19,20]. In practice, one can either use a pulsed laser to inject a short pulse of light or, as we do in this work, one can build up light energy in the cavity by illuminating it with a continuous-wave (cw) laser and subsequently extinguishing the laser to yield an exponential decay. A detector placed behind the cavity measures the intensity of light exiting the cavity, which also decays exponentially, yielding the ring-down signal. The exponential decay time constant (also termed ring-down time), $\tau$, is related to the total loss inside the cavity, which is due to the single-pass empty cavity loss $L_c$, and sample absorbance $Abs(v)$:

$$\frac{1}{\tau(v)} = \frac{c}{l} \cdot (L_c + Abs(v)) = \frac{1}{\tau_0} + \frac{c}{l}\int_0^d k(x,v)dx \quad (1)$$

where $c$ is the speed of light, $v$ is the laser frequency, $l$ and $d$ are the cavity length and the sample length respectively, $x$ is the position along the optical axis, $k(x,v)$ is the absorption coefficient, and $\tau_0$ is empty cavity ring-down time (often measured by detuning the laser fron the sample absorption). The empty cavity loss $L_c$ is generally dominated by mirror transmission loss and for a single-pass is then equal to $1-R$. For concentration measurement as we perform, the goal is to measure the sample absorbance which follows from the ring-down time which is found by fitting the measured with an exponential.

A commonly used approach is to scan the laser frequency across the absorption line and to measure the wavelength- (or frequency-) integrated spectrum i.e. the line area. Assuming the spectroscopic line parameters are known, the measured area $\int Abs(v)dv$ of a transition from lower state $i$ to upper state $k$ can be readily converted to the path-integrated concentration of the lower state $\int N_i dx$ as:

$$\int_0^d N_i dx = 8\pi \frac{g_i}{g_k} \frac{v_{ki}^2}{A_{ki}c^2}\left(\int Abs(v)dv\right) \quad (3)$$

where $g_i$, $g_k$ are the level degeneracies, $v_{ki}$ is the transition frequency, and $A_{ki}$ is the transition Einstein $A$ coefficient. For cases where the spatial distribution of particles is non-uniform, actual concentration profiles can be determined from the path-integrated concentration in several ways. For rough approximation one can assume a uniform concentration profile over a known column length, $d$. Alternatively, Abel inversion or other inversion approaches based on inversion and modeled spatial profiles can be used [24].

*2.2. Boron Nitride Detection Scheme*

Our approach for detection of boron nitride (BN) is based upon CRDS absorption measurement of sputtered boron atoms. (Atomic nitrogen is not a readily optically accessible species.) The exact composition of the sputtered particles from BN, i.e. proportions of B, N, $B_xN_y$ etc., is not well understood and may vary with sputtering conditions. Sputter yield measurements indicate that the majority of sputtering is in the form of boron and nitrogen atoms, perhaps with a small fraction of BxNy clusters [29]. It is also possible that the sputtered nitrogen is in the form $N_2$ or $N_x$, but this does not affect our measurements. Based on these findings, we assume all sputtering is as atoms so that each sputtered boron atom corresponds to ejection of one boron atom and one nitrogen atom from the BN surface. This assumption can later be refined based on species-specific sputter yield measurements and modeling [30].

A partial energy level diagram for neutral boron (B I) is shown in figure 1. The ground term has two distinct levels: $2s^2P_{1/2}^0$ (0 eV) and $2s^2P_{3/2}^0$ (0.00189 eV). As a result, fine-structure splitting results in two distinct B I absorption lines near 250 nm: the $2s^2P_{1/2}^0 \rightarrow 3s^2S_{1/2}$ transition at 249.753 nm, and the $2s^2P_{3/2}^0 \rightarrow 3s^2S_{1/2}$ transition at 249.848 nm. These lines are selected for CRDS measurement based on their optical accessibility and high absorption strength. From the energy level information of boron atoms and partition function, more than 99.9% of all boron atom's population is calculated to reside in the split ground state (66.6% of all population in $2s^2 P^0_{3/2}$). Therefore, to a good approximation all population will reside in the split ground state, meaning that the measured states will provide a direct measure of the overall boron population. (This is contract to OES approaches that probe a small fraction of the sputtered particles residing in upper energy-levels.)

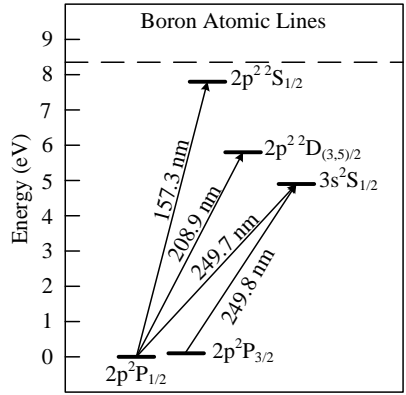

**Figure 1.** Partial energy level diagram of atomic boron

## 3. Experimental

*3.1. Diagnostic Apparatus*

For development and testing of the CRDS sensor, a diagnostic testbed based on an ion source is used. The gridded ion source provides a collimated mono-energetic beam that facilitates measurement interpretation. The system comprises a sputtering apparatus within a vacuum chamber containing an optical rail system. Figure 2 shows a schematic diagram of the diagnostic apparatus and sensor setup. With roughing and turbo pumps (Turbo-V550), the pressure inside the vacuum chamber reaches approximately $10^{-4}$-$10^{-5}$ Pa under no-flow conditions and $10^{-4}$ Pa under a small feeding argon flow (~0.18 mg/s) for the ion source. At these conditions, the sputtered atoms are in a free-molecular regime (Knudsen number << 1). The ion beam is extracted from an 8 cm diameter two-grid ion source using refractory metal filaments for both the main and neutralizer cathodes [31]. The ion source operates with an commercial power supply (IonTech MPS 3000), with typical beam currents and voltages of about 10–100 mA and 400–1200 V respectively. The target, currently a piece of BN (15 cm long in the beam direction, 12 cm wide), is mounted on a stainless steel post and is (approximately) normal to the ion beam. It is located 20 cm away from the ion source exit and 1 cm from the optical axis. The BN sample is HBC grade from General Electric's Advanced Ceramics. Due to the large divergence of the ion beam from our ion source, 12% of the beam current is measured to hit the target.

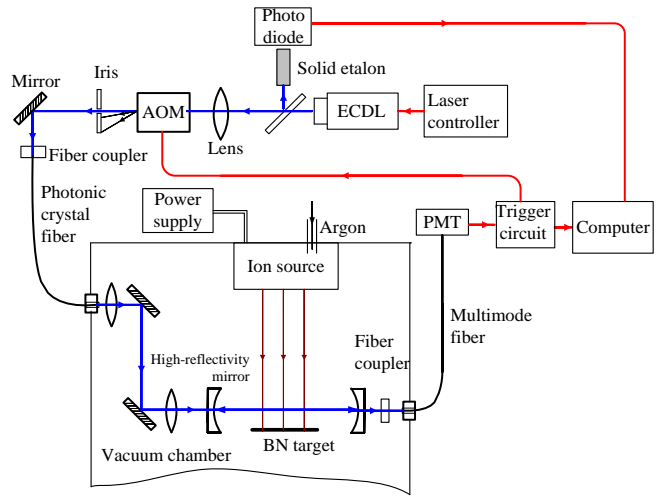

**Figure 2.** Schematic diagram of diagnostic apparatus

A custom designed optical rail system is used to hold the CRDS optics inside the vacuum chamber. As shown in figure 3, the rail system includes aluminum end plates connected by carbon fiber rods and mounted to a base plate. The carbon fiber rods have a nearly zero coefficient of thermal expansion in order to minimize optical misalignment due to the rods deflecting under from heat loading of the Hall thruster (or ion source). The laser beam is brought in and out of the vacuum chamber with fiber optic cables (see below). A 30 mm optical cage system is used to mount all optics, including fiber connector, alignment optics, high reflectivity mirrors. In order to protect the mirrors from sputter deposition, two 10 cm long tubes containing a series of 2 mm diameter irises inside are mounted to the end plates in front of each mirror. This allows us to have a longer operation time (tens of hours) with minimal degradation of mirror reflectivity and detection sensitivity. Four rubber damping feet are used to minimize the effect of vibrations from the pump system. Custom made feed-throughs are used for the optical fibers.

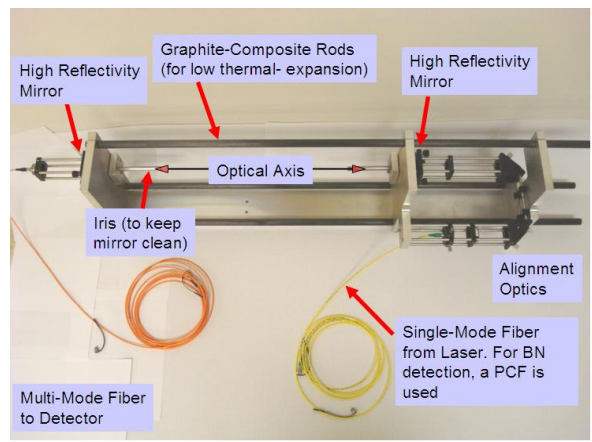

**Figure 3.** Fiber-coupled rail for implementation within vacuum chamber

The advantage of such a modular design is that experiments can be conducted in various vacuum facilities with little or no modification. Moreover, the rail system allows for the optical setup to be tested and adjusted outside of the vacuum chamber, and maintains alignment as the chamber is pumped down. This is in contrast to our past efforts [21-24] in which the cavity mirrors were mounted on the outer walls of the vacuum chamber and which suffered from alignment degradation as the chamber was pumped down or heated (and the walls deflected).

*3.2. CRDS Setup*

The elements of the cw-CRDS setup are also shown in figure 2. A frequency-quadrupled external cavity diode laser (ECDL) system (Toptica TA-FHG110) with a linewidth less than 5 MHz is used as the continuous-wave light source. It can generate a 60 GHz mode-hop free tuning range around the targeted transition lines of boron, with an output power of ~10 mW. The diode laser system was tuned to scan mode-hop free around the 249.848 nm, which is the stronger absorption line. One scan takes about 12 s (up and down). A beam splitter is placed in front of the laser and sends about 10% of the laser power through a solid Fabry–Pérot etalon (finesse of 10 at 250 nm) onto a photodiode to generate a frequency reference signal. The rest of the laser beam goes though an acousto-optic modulator (AOM) with the first order output beam (~70% of the input laser power) going to the ring-down cavity. The AOM is used as a fast optical switch activated by a trigger circuit to extinguish the laser beam and produce the ring-down signal. The beam is coupled into the vacuum chamber using a 4 m single-mode photonic-crystal fiber with a core diameter of ~10$\mu$m (Crystal Fibre LMA-10 UV). Details of the PCF delivery are given in Section 3.3. At the output of the PCF, two plane-convex fused silica lenses ($f$ =20 mm, $f$ =500 mm) are used to mode match the laser beam to the fundamental longitudinal mode ($TEM_{00}$) of the cavity. The optical cavity has length of 75 cm and is formed by two high reflectivity mirrors each 2.54 cm in diameter with 1 m radius of curvature. An empty-cavity ring-down time of around 1.3 $\mu$s is achieved, corresponding to $R$=99.8%. The reflectivity is low relative to that achievable in the visible and near-infrared, but is reasonable for dielectric layer mirrors in this spectral region. Behind the cavity, the output beam is coupled into an ultraviolet multimode fiber that delivers light to a photomultiplier tube (PMT model Hamamatsu R9110) with a dielectric interference filter to suppress background light. Ring-down signals are measured each time the laser spectrally overlaps a cavity transmission peak during the wavelength scans of ECDL. The trigger circuit causes the AOM to extinguish the light, resulting in a ring-down event measured by the PMT. A computer with a 20 MHz, 12 bit analog-to-digital acquisition board (Adlink PCI-9812) collects ring-down signals when triggered by the trigger circuit. A LABVIEW program is used for laser frequency calibration and exponential fitting (nonlinear Levenberg-Marquartdt fit).

*3.3. Fiber Delivery with Photonic Crystal Fibers*

Fiber delivery provides great flexibility in handling the laser light in harsh environments. For the CRDS erosion sensor, delivery of 250 nm UV light into the vacuum chamber by optical fiber is strongly desirable, as it avoids many complications associated with cavity mirrors mounted on the wall and delivery mirrors external to the vacuum chamber [21-24]. Certain fiber optics also allows single-mode output as is required for efficient mode match and cavity coupling. However, for conventional single-mode silica step-index fibers, the needed core diameter is ~2 $\mu$m for delivery of 250 nm light. Such small core diameters make efficient fiber input coupling very difficult [32]. We have recently shown that photonic crystal fibers

(PCFs) can be used for single-mode delivery at the needed UV wavelengths [33]. Typical PCFs have a uniform patterned microstructure of holes running axially along the fiber channel with a missing hole in the center providing a core region. With appropriate fiber design, the fiber core can support a single guided mode over all optical frequencies, a characteristic referred to as "endless single-mode operation."

In the CRDS setup, a large mode area PCF (Crystal Fibre LMA-10 UV) with a core diameter of ~10 μm was employed and mounted in a V-groove fiber holder on a precision 5-axis stage. The fiber is optimized for ultraviolet and visible operation [34]. Figure 4 shows a photograph of the fluorescence light caused by the 249 nm UV laser output on a card several cm downstream of the fiber exit. The shape of the beam pattern is typical of single-mode output from a triangular PCF [35]. Single mode output with reasonable transmission efficiency (~ 45% for 1 m PCF) over short distances (<10 m) has been obtained. Both launch loss and transmission loss have been examined [33]. An issue to consider is that optically induced damage of silica by UV photons was observed [36,37]. Depending on laser power, the time of damage (manifested as reduced transmission) is around one to tens of hours. For input power of 0.3 mW, the damage is reduced with a decrease in transmission of <40% over more than 40 hours of operation [33]. Clearly such optical damage is challenging in long-term applications (e.g. communication links), but the PCFs do allow sufficient operation times for our cw-CRDS BN sensor where typical measurement campaigns to study several thruster operating conditions require approximately one to several hours.

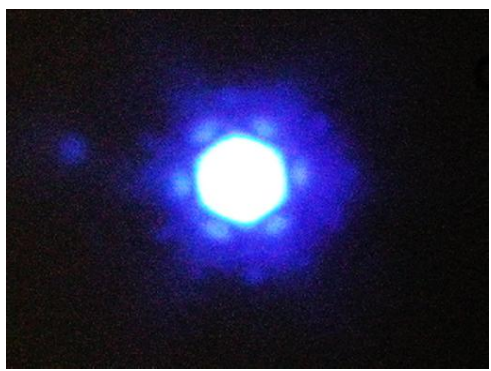

**Figure 4.** Photograph of single-mode output beam from PCF

4. Results and Discussion

*4.1. Demonstrative Measurement and Validation*

An example of boron spectrum measured by the CRDS sensor is shown in figure 5. The spectrum was measured in the diagnostic testbed and is due to argon ions of current 50 mA and energy of 1200 eV sputtering the BN target in 20 s collection time. The absorbance spectrum is plotted in units of parts per million (ppm). (Note that the ppm refers to the optical absorbance not the boron concentration.) To construct the spectrum, a binning approach dividing the frequency axis into a series of 1 GHz-wide bins, was used. Within each bin, the experiment data points were averaged both by frequency and ring-down time. A Voigt profile was fitted to the experiment data points. Path-integrated concentrations were calculated from equation 3 using the wavelength-integrated area of the spectrum. The true lineshape is not a single Voigt lineshape, but a Voigt lineshape is used as a convenient means to fit data and determine the area. In some cases, we have also numerically integrated to find area and the consistency of the two methods is within our experimental error.

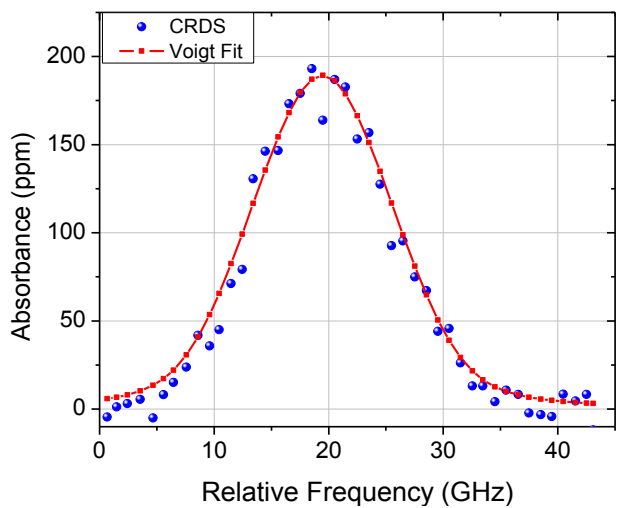

**Figure 5.** Sample boron absorption spectrum.

Tests were made to validate our cw-CRDS sensor results. Measurements of boron number density were made as the beam current and voltage were varied. Figure 6(a) shows a plot of the dependence of path-integrated boron number density on the ion voltage. The ion beam voltage was kept at 1200V, while the ion beam current varied from 30 mA to 60 mA. The sputter yield per incident ion does not change with beam current density. Therefore, the path-integrated number density of boron is expected to be proportional to the beam current and the correlation coefficient $R$ is 99.8% from linear fitting. Similarly, figure 6(b) shows a plot of dependence of path-integrated boron number density on the ion beam voltage. The sputter yields per incident ion changes almost linearly in this voltage range and again the path-integrated number density shows good linearity with a correlation coefficient of 99.6%. These results contribute to validation of the basic performance of the cw-CRDS sensor.

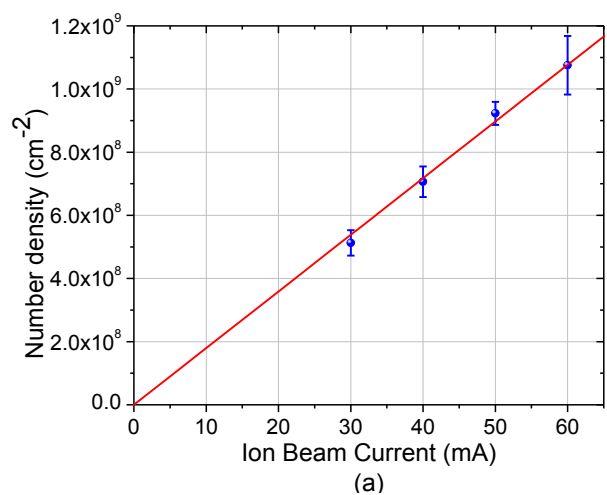

(a)

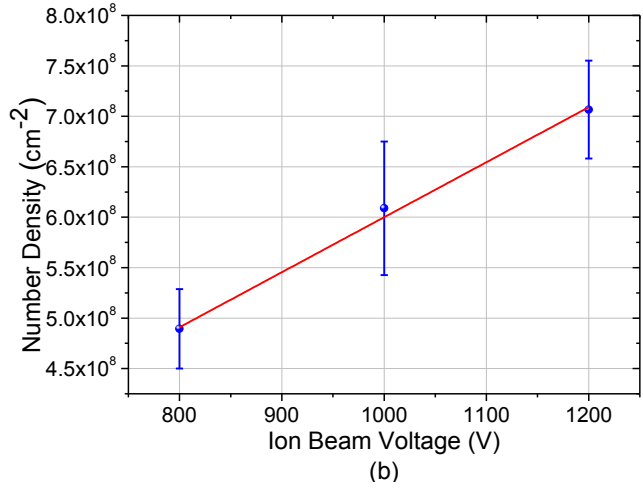

(b)

**Figure 6.** (a) Dependence of path-integrated boron number density on the ion beam current at $V_b$ = 1200V. b) Dependence of path-integrated boron number density on the ion beam voltage at $I_b$ = 40mA.

As described in our previous work [21,23], a mathematical model for analyzing the CRDS sputtering signals for known ion beams and simple geometry has been developed with a simple finite element approach. Table 1 shows the comparison between CRDS measured number density and the calculated number density from our finite element model, for the case of beam voltage of 1200V and varying beam current. We assume a uniform current density distribution on the target, a diffuse shaped sputter yield, and adopt the binding energy $E_b$ = 3.865 eV ($v_b$ = 8300 m/s) with an exponent of 1.8 in the Thomson distribution for particle velocity [27]. The sputter yield for BN at 1200 V is interpolated from measured yields [29], and a diffuse cosine profile is assumed. The uncertainty on experimental values is from Einstein A coefficient and reproducibility of the measurements, while the uncertainty in the model (± 40%) has contributions from total sputter yield (± 15%) [29], shape of differential sputter yield (± 5%), velocity distribution (± 30%), and current density distribution (± 12%). We find favorable agreement (within error bars) between the model and experiment, providing further validation on the sensor. Future efforts include experimental measurement of the sputtered particle velocity distribution, both by CRDS [22] and Laser induced fluorescence. The latter technique has been extensively demonstrated for velocity measurements of sputtered particles [10-11]. Knowledge of particle velocity is also needed to relate CRDS number densities to particles fluxes, as is needed to infer thruster erosion rates from the CRDS results.

**Table 1.** Measured and modeled boron number density for $V_b$ = 1200V

| Beam current (mA) | CRDS results (cm$^{-2}$) | Mathematical model results (cm$^{-2}$) |
| --- | --- | --- |
| 30 | $7.7 \pm 0.4 \times 10^8$ | $9 \pm 4 \times 10^8$ |
| 40 | $10.7 \pm 0.5 \times 10^8$ | $12 \pm 5 \times 10^8$ |
| 50 | $13.9 \pm 0.5 \times 10^8$ | $15 \pm 7 \times 10^8$ |
| 60 | $16.3 \pm 0.9 \times 10^8$ | $18 \pm 8 \times 10^8$ |

*4.2. Sensitivity Analysis*

Estimates of minimum detectable optical absorbance can be found from the noise in the cw-CRDS detection system. The minimum detectable absorbance, $Abs_{Min}$, is calculated from

$$Abs_{Min} \equiv \frac{\Delta\tau}{\tau}(1-R) \quad (4)$$

where $\Delta\tau$ is the uncertainty (noise) in measurement of ring-down time $\tau$, for which we use the standard error in measurement of $\tau$. The standard error represents the expected standard deviation between the measured estimates and the true value, and is computed as the standard deviation divided by the square-root of the number of measurements. The standard deviation of our cw-CRDS system is approximately 1% of the ring-down time $\tau$. For a fixed frequency, the minimum detectable absorbance is about 0.6 ppm in 20 s measurement time (~1000 ring-down events). In a typical frequency scan over 50 GHz, there are about 20 ring-down measurements within each 1 GHz bin in 20 s. The resulting minimum detectable absorbance due to the BN absorption line is about 5 ppm for 20 s measurement time.

To assess the expected signal-to-noise ratio (SNR) for thruster studies we consider, the expected boron concentration in the thruster plume. Based on scaling of wall erosion rates modeled by Yim *et al.* [38], a path-integrated number density of boron atoms at the thruster exit plane of $10^9$ cm$^{-2}$ to $10^{10}$ cm$^{-2}$ was predicted. The corresponding peak absorbance of the boron absorption line at 249.848 nm is estimated to be ~30-300 ppm (based on an expected full-width half maximum of 20-50 GHz). Thus, for expected thruster conditions the SNR of our CRDS boron sensor would be ≈6-60 for a 20 s measurement time, and can be increased with longer measurement durations. It is also interesting that, especially for long measurement times, the detection limit of our CRDS system is sufficiently low, that if the sensor does not yield a measurable signal, then we will have confirmed that the boron concentrations (~$10^6$ cm$^{-3}$) and associated erosion rates (~0.1 μm/hr) are low enough to correspond to very long thruster lifetimes (>9,000 hours).

## 5. Conclusion

Techniques providing in situ quantitative measurements of Hall thruster erosion rates are of importance for predicting the lifetime and for optimizing operation conditions and thrusters designs. The CRDS sensor presented here should contribute to filling the current gap in the ability to rapidly measure BN erosion. In this work, a BN erosion sensor has been developed based on CRDS using a frequency-quadrupled ECDL in the vicinity of the boron atomic transition (~ 250 nm). The optical cavity is housed on a rail system with fiber coupling in and out of the vacuum chamber. A PCF was used to deliver single-mode UV laser light into the cavity. The flexibility of such an approach allows implementation in different types of vacuum chamber facilities.

We have measure the absorption spectrum of boron atoms sputtered by an ion beam from a BN target sample as an initial demonstration of the sensor. The dependence of sputtered particle number density on ion beam current and voltage was measured and found to be in accord with expectations. The measurements were also validated against finite-element sputter model. The sensor has a minimum detectable absorbance of 0.6 ppm in 20 s measurement time, which should be very adequate for erosion studies of Hall thrusters. The sensitivity is largely limited by the relatively poor reflectivity of the ultraviolet mirrors. One avenue for improvement may be to replace these with $CaF_2$ prism-based retro-reflectors [39,40].

**Acknowledgments**

The authors would like to acknowledge the assistance of Brian Lee and Prof. John Williams at Colorado State University, and Dr. Tim Smith at University of Michigan. This work was supported in part by an AFOSRDURIP grant (FA9550-07-0535). Dr. Mitat Birkan is the Project Monitor for this grant.